\documentclass[prd,amsmath,amssymb,showpacs,showkeys,floatfix,
               nofootinbib,eqsecnum,
               preprint,12pt,tightenlines,fleqn
               ]{revtex4-1}
\usepackage{graphicx}

\def\al{\alpha}

\def\ga{\gamma}
\def\de{\delta}

\def\et{\eta}

\def\ka{\kappa}
\def\la{\lambda}

\def\rh{\rho}

\def\si{\sigma}

\def\om{\omega}
\def\Ga{\Gamma}

\def\mn{{\mu\nu}}

\def\prt{\partial}

\newcommand{\beq}{\begin{equation}}
\newcommand{\eeq}{\end{equation}}
\newcommand{\beqa}{\begin{eqnarray}}
\newcommand{\eeqa}{\end{eqnarray}}
\newcommand{\bse}{\begin{subequations}}
\newcommand{\ese}{\end{subequations}}

\newcommand{\BB}{\big}

\newcommand{\W}{\widetilde}
\newcommand{\x}{\times}

\usepackage{amsmath}
\usepackage{amsfonts}
\usepackage{amssymb}

\newcommand{\bM}{\begin{pmatrix}}
\newcommand{\eM}{\end{pmatrix}}

\newcommand{\LVga}{\widetilde\ga}

\begin{document}

\noindent
Phys. Rev. D 94, 085025 (2016) \hfill arXiv:1607.02099\newline\vspace*{4mm}
%
%

\title{Changes in extensive air showers from isotropic Lorentz violation in the photon sector}
\author{J.S. D\'iaz}
\email{jorge.diaz@kit.edu}
\author{F.R. Klinkhamer}
\email{frans.klinkhamer@kit.edu}
\affiliation{Institute for
Theoretical Physics, Karlsruhe Institute of
Technology (KIT), 76128 Karlsruhe, Germany}
\author{M. Risse}
\email{risse@hep.physik.uni-siegen.de}
\affiliation{Physics Department, University of Siegen, 57068 Siegen, 
Germany\vspace*{10mm}}

\begin{abstract}
\noindent
\vspace*{-4mm}\newline
We consider a theory with isotropic nonbirefringent Lorentz violation in the photon
sector and explore the effects on the development of the electromagnetic component of extensive air showers in the Earth atmosphere.
Specifically,
we consider the case of a ``fast'' photon with a phase velocity
larger than the maximum attainable velocity of a massive Dirac
fermion (this case corresponds to a negative
Lorentz-violating parameter $\kappa$ in the action).
Shower photons with above-threshold energies
decay promptly into electron-positron pairs,
instead of decaying by the
conventional production of electron-positron pairs in the
background fields of atomic nuclei.
This rapid production of charged leptons accelerates the shower development,
decreasing the atmospheric depth of the shower maximum ($X_\text{max}$)
by an amount which could be measured by cosmic-ray observatories.
Precise measurements of $X_\text{max}$ could then improve existing limits on the negative Lorentz-violating parameter $\kappa$
by several orders of magnitude.
\end{abstract}


\maketitle
\section{Introduction}

Ever since the foundation of special relativity,
electrodynamics has played a fundamental role in the establishment of
elementary-particle-physics theory.
Experimental tests using photons have provided valuable and sensitive probes for the search of potential deviations from exact Lorentz symmetry.
In fact, some of the most stringent limits on dimensionless parameters controlling Lorentz violation have been found in the photon sector.
Although laboratory experiments have access to some unique signatures,
the determination of most of the best limits on Lorentz-violating (LV)
parameters in various sectors have taken advantage of the high energies
or the large propagation distances of ``astroparticles,''
i.e., cosmic rays~\cite{KlinkhamerRisse1,KlinkhamerRisse2,KS2008,DK2015}, 
gamma rays \cite{KS2008}, cosmic-microwave-background photons \cite{KM2007},
and neutrinos \cite{DKM2014,Diaz:2016c}.

Here, we explore the potential of ultrahigh-energy photons
to test LV effects and consider photons which are produced as secondary particles in air showers. In Sec.~\ref{sec:Theory}, we give the
theoretical setup for isotropic Lorentz violation in the photon sector
and review existing bounds on the relevant LV parameter.
In Sec.~\ref{sec:EAS}, we discuss a simple model for the electromagnetic
component of extensive air showers, first for the standard Lorentz-invariant
theory and, then, for the  Lorentz-violating theory considered.
In Sec.~\ref{sec:summary}, we present some concluding remarks.

As the
scope of this article is restricted to the study of isotropic
Lorentz violation in electrodynamics,
we only consider isotropic Lorentz-violating effects
in the photon sector. Notice that the corresponding effects from the electron sector can be moved to the photon sector by suitable
spacetime coordinate transformations~\cite{BaileyKostelecky2004}.
Furthermore, independent studies regarding Lorentz-violating effects
involving weakly interacting particles
can be found for muons \cite{Noordmans-etal2014},
pions~\cite{Antonov-etal2001,Altschul2007,Boncioli-etal2015},
and neutrinos \cite{Diaz2014}.

\section{Theory}
\label{sec:Theory}

A relatively simple extension of standard
quantum electrodynamics~\cite{Heitler1944}
adds a single LV term~\cite{ChadhaNielsen1983,KM2002}
to the Lagrangian density. This term
breaks indeed Lorentz invariance but preserves CPT and gauge invariance.
The total Lagrangian density is then
\beq\label{L}
\mathcal{L} = -\frac{1}{4} F^{\mu\nu}F_{\mn}
+ \overline{\psi}\,\BB[\ga^\mu(i\prt_\mu-e\,A_\mu)-m\BB]\,\psi
-\frac{1}{4}(k_F)_{\mu\nu\rh\si} F^{\mu\nu}F^{\rh\si}\,,
\eeq
where the first term gives the standard propagation of the photon,
with Maxwell field strength $F_{\mu\nu}\equiv\prt_\mu A_\nu-\prt_\nu A_\mu$,
the second term gives  the standard propagation of a Dirac fermion (mass $m$ and electric charge $e$) and its standard minimal interaction with the photon field $A_\mu$, and the last term modifies the photon propagation
by adding a dimension-four operator for CPT-even Lorentz
violation~\cite{ChadhaNielsen1983,KM2002}.
The Minkowski metric
$g_\mn(x)=\et_\mn\equiv [\text{diag}(+1,\,-1,\,-1,\,-1)]_\mn$ and natural units with $\hbar=1=c$ are used throughout this article (see below for
a remark on the meaning of the velocity $c$).

The constant fixed ``tensor'' $(k_F)_{\mu\nu\rh\si}$ has 20 independent components,
ten of which produce birefringence and eight of which lead to direction-dependent modifications of the photon-propagation properties.
The remaining two components correspond to an isotropic modification of the photon propagation and an unobservable double trace that changes the normalization of the photon field.
The isotropic nonbirefringent violation of Lorentz invariance in the photon sector is then controlled by a single dimensionless parameter $\ka$, which
enters the fixed tensor $k_F$ of \eqref{L} in the following way:
\beq\label{k_F}
(k_F)^\la_{\phantom{\la}\mu\la\nu} = \frac{\ka}{2}\;
\Big[\,\text{diag}\big(3,\,1,\,1,\,1\big)\,\Big]_\mn
\,.
\eeq
Notice that the \textit{Ansatz} \eqref{k_F} gives $(k_F)^{\mn}_{\phantom{\mn}\mn}=0$,
which removes the unobservable double trace.
In physical terms, the velocity $c$ corresponds to the
maximum attainable velocity of the massive Dirac
fermion of \eqref{L}, whereas the phase velocity
of the photon is given by
\beq\label{v_ph}
v_\text{ph} =
\frac{\omega}{|\vec{k}|}=
\sqrt{\frac{1-\ka}{1+\ka}}\;c\,,
\eeq
which is smaller/larger than $c$ for positive/negative values of
$\ka$.

As mentioned in the Introduction,
isotropic Lorentz-violating effects can also
appear in the fermion sector. But, for a single fermion,
isotropic $c$-type Lorentz-violating effects can be moved to the photon
sector by a suitable change of spacetime coordinates \cite{BaileyKostelecky2004} (see also App.~B of Ref.~\cite{KS2008}).

For nonzero values of the LV parameter $\ka$
in the theory \eqref{L}--\eqref{k_F},
certain processes conventionally forbidden become allowed.
For positive $\ka$, a Cherenkov-type radiative process
can already occur in the vacuum: $f^\pm\to f^\pm+\LVga$, with
$f^\pm$ standing for an electrically-charged Dirac (anti-)fermion  and
$\LVga$ denoting the nonstandard photon of the theory considered.
Since such vacuum-Cherenkov processes remain unobserved so far,
constraints on the positive value of this parameter have been determined,
for example, by considering the physics of particle accelerators \cite{Hohensee2009,Altschul2009}.
Even better,
cosmic-ray data have reduced the upper limit to \cite{KS2008}
\beq\label{KS-upper-limit}
\ka < 6\times10^{-20}\quad\text{(98\% C.L.)}.
\eeq

For negative $\ka$ in the theory \eqref{L}--\eqref{k_F},
photons become unstable above the energy threshold
\beq\label{omega_th}
\om_\text{th} = 2\,m\;\sqrt{\frac{1-\ka}{-2\,\ka}}
\eeq
and decay into pairs of electrically charged fermions.
Hereafter,
we consider the Dirac fermion field $\psi$ to be the electron/positron field
with the corresponding constants
\beq\label{m-e-electron}
(m,\,e) = (m_\text{electron},\,e_\text{electron})\,.
\eeq
The tree-level photon decay (PhD) rate as a function of the photon energy
$\om \geq \om_\text{th}$ is given by~\cite{KS2008,DK2015}
\beq\label{eq:Gamma-PhD}
\Ga_\text{PhD}(\om) = \frac{\al}{3}\,\frac{-\ka}{1-\ka^2}\, \sqrt{\om^2-\om^2_\text{th}} \; \Big(2+\om^2_\text{th}/\om^2\Big),
\eeq
where $\al \equiv e^2/(4\pi) \approx 1/137$  is the fine-structure constant 
and $m \approx 511\:\text{keV}$ is the electron mass entering the
threshold energy $\om_\text{th}$ from \eqref{omega_th}.   

As photons above threshold decay very efficiently into pairs
($\LVga\to e^-+e^+$),
only below-threshold photons will reach Earth from distant sources
and will travel through the Earth atmosphere. For this reason,
the observation of energetic gamma rays in the Earth atmosphere can be
used to determine a bound on negative $\ka$. In fact,
the most stringent lower limit has been obtained from the terrestrial
observation of gamma rays \cite{KS2008}
\beq\label{KS-lower-limit}
-9\times10^{-16}<\ka \quad\text{(98\% C.L.)}.
\eeq
For completeness, we mention that an astrophysics
lower bound of order $-6 \times 10^{-20}$
has been obtained by considering synchrotron radiation in the
Crab Nebula~\cite{Altschul2005} (the $\ka$ bound quoted  
is from Ref.~\cite{DK2015}). This astrophysics  bound is,             
however, only qualitative, different from  
bound \eqref{KS-lower-limit} with a specified confidence level.
Moreover, bound \eqref{KS-lower-limit}
does not rely on assumptions about the astronomical source,
the only input for the energy determination of the gamma ray
being well-known physical processes operating in the environment
of the Earth atmosphere.

Since the lower limit \eqref{KS-lower-limit} from terrestrial
observations of gamma-ray primaries
is about four orders of magnitude weaker than the upper limit
\eqref{KS-upper-limit} from cosmic-ray primaries,
we would like to find a process that could allow further improvements
and that would be accessible to terrestrial observations
(e.g., in the Earth atmosphere).

\section{Extensive air showers}
\label{sec:EAS}

\subsection{Preliminary remarks}
\label{subsec:Preliminary-remarks}

We now present a new method to explore the negative range of values
of the LV parameter $\ka$ as defined in Sec.~\ref{sec:Theory}.
Even though gamma rays have already served as tools to study this part of parameter space, we present a novel method,
in which above-threshold photons can be produced in particle cascades in the Earth atmosphere. We will show that
Lorentz-violating effects for sufficiently large
(but numerically still very small)
values of $\ka$
can drastically modify the development of the electromagnetic components of the shower.

When a cosmic-ray hadronic primary impacts the upper Earth atmosphere,
charged and neutral pions are produced.
The high-energy charged mesons usually continue to feed the hadronic shower, while the neutral pions rapidly decay into photons triggering an electromagnetic shower.
The particle cascades produced by the interaction of the primary cosmic ray with the atmosphere are known as extensive air showers (EAS).

Since we are ultimately interested in the effects of modified electrodynamics,
we focus on the development of the electromagnetic shower.
Consider, first, the standard Lorentz-invariant
theory (LV parameter $\ka=0$).
After two photons of energy $\om_{0}$ are produced from the decay
of neutral pions ($\pi^0$),
their propagation in the Coulomb fields
of the atomic nuclei of the atmosphere
will lead to the production of electron-positron pairs,
the ``external conversion'' process. Subsequently,
each charged lepton will produce new photons by 
the Bremsstrahlung process.
These are the two relevant high-energy
processes for the standard theory and they
give an exponential multiplication of the number of particles in the electromagnetic shower.

\subsection{Heitler model}
\label{subsec:Heitler-model}

A detailed description of the shower development requires a
numerical simulation of the particle production as well as a probabilistic distribution of energy between the daughter particles at each interaction point. Still,
a general analytic description of the main properties of the
electromagnetic shower can be obtained by use of the Heitler model
(see  Sec.~24 of Ref.~\cite{Heitler1944} and references therein),
a simplified representation of the processes that produce new photons and charged leptons in the shower.
This simple model accurately describes the most important features of the electromagnetic component of the shower.

The Heitler model depicts the particle multiplication
as a binary tree (cf. Fig.~27 of Ref.~\cite{Heitler1944}
and Fig.~2 of Ref.~\cite{UHECRreview2011}):
\begin{itemize}
  \item each photon produces two charged leptons via pair production
  in the presence of a nucleus,
  \item each new charged lepton produces a lepton and a photon via Bremsstrahlung.
\end{itemize}
Consider an initial photon of energy $\om_{0}$.
After $n$ generations, the shower will then contain $2^n$ particles
and the energy of the parent particle will be split in equal parts,
each particle having an energy 
\beq
E_n=\om_{0}/2^n\,.
\eeq
The particle multiplication will continue until the individual particle energy reaches a critical value $E_c$ and
energy loss by ionization starts to dominate over radiative processes.
In air, the critical value $E_c$ is of order \cite{UHECRreview2011}
\beq\label{E_c}
E_c \approx 80\; \text{MeV}.
\eeq
Using the similarity of the pair-production and Bremsstrahlung cross sections,  
the interaction step length $d$
of charged leptons and photons can be approximated as being equal. Hence,
the shower reaches its maximum after propagating in the Earth atmosphere to a depth
\beq\label{X_max}
X_\text{max} = \lambda_{r}\, \ln(\om_{0}/E_c),
\eeq
where $\lambda_{r} \equiv d/\ln2$
is the radiation length in the medium.
For air, we have \cite{UHECRreview2011}
\beq
\lambda_{r} \approx 37$ g/cm$^2.
\eeq
The depth of the shower maximum
\eqref{X_max} is a quantity that cosmic-ray observatories
such as the Pierre Auger Observatory~\cite{Auger2015}
can measure reliably.
For $10^{19}\:\text{eV}$ showers observed by Auger, the systematic uncertainty in $X_\text{max}$ is approximately $10\,\text{g/cm}^2$ and the resolution is approximately $15\,\text{g/cm}^2$~\cite{Auger2014}.

\subsection{Shower model with Lorentz violation}
\label{subsec:Shower-model-with-LV}

Consider, next, the effects from the Lorentz-violating modification
in \eqref{L} on the development of the electromagnetic shower.
In the presence of isotropic nonbirefringent Lorentz-violating electrodynamics \eqref{k_F},
the development of the shower will be modified by nonstandard
decay processes which become allowed.
For a negative value of the LV parameter $\ka$, in particular,
photons above the threshold energy \eqref{omega_th}
become unstable and rapidly decay into electron-positron pairs.
The expression for the decay rate has already been given in
\eqref{eq:Gamma-PhD}.

First, take the LV parameter $\ka$ to have the numerical
value $-9\times10^{-16}$,
which is consistent with the quantitative bound \eqref{KS-lower-limit}.
Figure~\ref{Fig:l_PhD} shows the photon decay length $l_\text{PhD}\equiv 1/\Ga_\text{PhD}$ in meters for this value of $\ka$
(the result for a $10^4$ times
smaller value of $\ka$ is also shown and will be discussed later).
The figure makes clear that photon decay is a very efficient energy-loss mechanism: above-threshold photons from astrophysical sources
cannot reach Earth as they promptly decay into charged leptons.
The corresponding threshold energy for photon decay into an electron-positron pair is $\om_\text{th}\approx24$ TeV, as follows from \eqref{omega_th}
with $\ka= -9\times10^{-16}$ and
$m = m_\text{electron} \approx 511\:\text{keV}$.

Very energetic cosmic-ray hadronic primaries can have energies several orders of magnitude above this 24 TeV threshold.
If about 10\% of the primary energy gets passed along to the first photons in the EAS through neutral-pion decay
(see App.~\ref{app:Heuristics}
for further remarks on this decay process in the LV theory),
these photons initiating the electromagnetic cascade will be above threshold.
The standard characteristic radiation distance
for pair production in air is of the order of kilometers,
whereas Fig.~\ref{Fig:l_PhD} shows that a photon
with an energy above threshold will decay after a few millimeters or less
(or after some 10 cm or less
for a $10^4$ times smaller value of $\ka$).
The subsequent electromagnetic shower would be drastically modified because these photons would promptly produce electron-positron pairs via photon decay rather than the conventional pair-production mechanism.
In the following generations,
a charged lepton produces a Bremsstrahlung photon with energy $\om>\om_\text{th}$, which will
almost immediately decay into a new pair of charged leptons.
We can call these  ``PhD-leptons,''
in order to distinguish them from the charged leptons
(called ``Br-leptons'')
which are produced with the Bremsstrahlung emission of photons.

Inspired by the Heitler model described in Sec. \ref{subsec:Heitler-model},
we follow
a similar approach to present a general description of the main features of the modified shower.
In our case, the shower proceeds as follows,
with $\LVga$ denoting the nonstandard photon described by the
theory \eqref{L}--\eqref{k_F}
with negative LV parameter $\ka$.

A $10^{15}\,\text{eV}$ -- $10^{20}\,\text{eV}$ cosmic-ray
hadronic primary impinges on the Earth atmosphere, producing charged and neutral pions.
Photons from subsequent neutral-pion decays ($\pi^0\to 2\,\LVga$) will then
have energies above the threshold \eqref{omega_th} and will promptly decay into electron-positron pairs $\LVga\to e^-+e^+$.
Each of these PhD-leptons ($e^\pm$) will produce a Br-lepton and a photon,
with the latter rapidly decaying into a new PhD-lepton pair. The
combined process is then given by
\beq\label{combined-process}
e^\pm\to e^\pm+\LVga\Rightarrow e^\pm+(e^-+e^+) \,,
\eeq
where the double arrow indicates the prompt nonstandard decay of $\LVga$.
Note that, for the small values of $\ka$ considered,
the rate of the
first (Bremsstrahlung) process in \eqref{combined-process} is essentially the same as in the Lorentz-invariant theory.

Since the photon decay in \eqref{combined-process}
occurs almost immediately after the Bremsstrahlung emission,
effectively three charged leptons are produced at each generation,
which share the energy of the parent particle.
Let us assume that, at each generation, three leptons are produced: one Br-lepton and two PhD-leptons, each having one-third of the total energy.
Notice that this is an approximate scenario because, even in an equal-energy-share situation,
the Br-lepton would take half of the parent's energy,
while the two PhD-leptons would each take one-fourth 
of the parent's energy. Still,
this very simplified model allows an analytic description of the shower that provides a general insight into the way the modified shower will develop.

A photon with energy $\om_{0}\gg\om_\text{th}$ rapidly decays into two PhD-leptons, each having an equal energy $\om_{0}/2$ by assumption.
Then,
each of these two leptons develops a branch whose number of particles grows as $3^{n'-1}$,
each with energy
\beq
E_{n'} = \frac{\om_{0}/2}{3^{n'-1}}\,.
\eeq
Since the first photon decay occurs immediately,
the parent photon with energy $\om_{0} \gg \om_\text{th}$
barely propagates through the atmosphere before decaying.
Hence,
the depth of the shower in the atmosphere is given by $X=(n'-1)\,d$.
The particle multiplication in the shower continues in this manner until the particles reach an energy below the threshold $\om_\text{th}$ for photon decay.
After this,
the shower continues in the conventional way as described in Sec.~\ref{subsec:Heitler-model}.
For $\om_{0}/2 > \om_\text{th} > E_c$,
the shower-maximum depth of this modified shower in the atmosphere can then
be written as the following sum:%
\beq\label{widetilde-Xmax}
\W X_\text{max} = \lambda_{r}\,\et\,\ln\left(\frac{\om_{0}/2}{\om_\text{th}}\right)  + \lambda_{r}\,\ln\left(\frac{\om_\text{th}}{E_c}\right) ,
\eeq
with
\beq
\et\equiv \ln2/\ln3\,.
\eeq
For $\om_{0} <  \om_\text{th}$,
the shower-maximum depth is given by the standard result \eqref{X_max}.
For intermediate values,
$\om_\text{th} \leq \om_{0} \leq 2\, \om_\text{th}$,
the function
$X_\text{max}(\om_{0})$ interpolates between
the standard behavior \eqref{X_max} and
the nonstandard behavior \eqref{widetilde-Xmax}.

The first term in \eqref{widetilde-Xmax} describes the part of the shower above the threshold energy \eqref{omega_th}
and the second term in \eqref{widetilde-Xmax} describes the part of the shower which begins with this threshold energy and ends when the energy per particle falls below $E_c$ given by \eqref{E_c}.
The factor $\et$ in the first term of \eqref{widetilde-Xmax}
traces back to the fact that each generation of the LV shower model
produces 3 particles, whereas each generation of
the standard Lorentz-invariant shower model of Sec.~\ref{subsec:Heitler-model} produces only 2 particles.

The calculated $\W X_\text{max}$ from \eqref{widetilde-Xmax} is shown in
Fig.~\ref{Fig:Xmax(w)} for two values of $\ka$ in comparison to the
standard $X_\text{max}$ from \eqref{X_max}.
We see a significant reduction of the depth of
shower maximum. For instance, for $\omega_0 = 10^9\:\text{GeV}$,
the depth of shower maximum is reduced by
some $160$\,g/cm$^2$ for $\kappa = -9\times 10^{-16}$
and some $100$\,g/cm$^2$ for $\kappa = -9\times 10^{-20}$.
Such reductions are large compared to the experimental uncertainties.
In addition, it is interesting to note that the factor $\eta$
in \eqref{widetilde-Xmax} diminishes the slope
(or ``elongation rate'') of the shower-maximum depth as a function
of $\ln \omega_0$.

The expression \eqref{widetilde-Xmax} for the modified shower-maximum depth
can also be written as a change of the conventional depth \eqref{X_max},
\beq\label{widetilde-Xmax-delta}
\W X_\text{max} =  (1+\de)\,X_\text{max}\,,
\eeq
with
\beqa\label{delta}
\de &=& \frac{ (\et-1)\ln(\om_{0}/\om_\text{th}) -\et\ln2}{\ln(\om_{0}/E_c)}\,.
\eeqa
Since $\om_{0}\gg\om_\text{th}$ by assumption and $\et<1$ by definition,
we find that $\de$ is negative, which implies that
the shower-maximum depth gets reduced.
This feature is to be expected because the particle multiplication
(and corresponding energy share) occurs faster
in the Lorentz-violating case than in the conventional case.

Figure~\ref{Fig:de(w)} shows this relative modification of
the shower-maximum depth for different values of the LV parameter $\ka$.
Expression \eqref{delta} relates the fractional depth change of the electromagnetic shower to the energy of the initiating photon $\om_{0}$.
This expression can be inverted for $\ka$ as a function of the photon energy $\om_{0}$, for a given precision $\de$ in the measurement of the
shower-maximum depth.
Figure~\ref{Fig:k(w)} shows the potential sensitivity to the LV parameter $\ka$ obtained from different values of the relative modification
$\delta$ of the shower-maximum depth.

Figures \ref{Fig:de(w)} and \ref{Fig:k(w)} show that the shower-maximum depth 
is particularly sensitive to a negative value of the Lorentz-violating 
parameter $\ka$.
A reliable estimate of the energy $\om_{0}$ of the first photon
and a careful determination of the allowed
deviation from the standard $X_\text{max}$ behavior
could improve the existing $\ka$ limit on the negative side 
by several orders of magnitude. For example,
the absence of deviations from the conventional maximum depth within 5\% 
for initial photon energies $\om_{0}$ of order $10^{7}\,\text{GeV}$  
would increase the excluded parameter space by four orders of magnitude.
In that case, the allowed range of negative values of the
LV parameter $\ka$ would be reduced significantly and would become
similar to the allowed range \eqref{KS-upper-limit}
of positive values obtained from the observation of energetic cosmic rays.

\section{Summary}
\label{sec:summary}

In this article,
we considered a theory with isotropic nonbirefringent Lorentz violation in the photon sector
and studied the effects on the development of extensive air showers.
In particular,
we focused on the consequences of a negative value for the
Lorentz-violating parameter $\ka$ because positive values are already well constrained \cite{KS2008}.

The relevant nonstandard process for $\ka<0$ is photon decay
into an electron-positron pair \cite{KS2008},
which substantially modifies the development of the electromagnetic component of the shower.
Since the photon decay length is of the order of a meter or less,
photons produced by the decay of neutral pions or by
subsequent Bremsstrahlung emission will promptly decay
into electron-positron pairs,
well before conventional pair production can take place.
This nonstandard decay process accelerates the particle multiplication
in the shower, which reaches its maximum earlier than for the conventional Lorentz-invariant case.

Two stages are identified in this electromagnetic shower.
In the first stage,
the rapid Lorentz-violating decay of Bremsstrahlung-photons produced with energies above the photon-decay
threshold will effectively lead to the production of three charged leptons at each generation, one ``Bremsstrahlung-lepton'' and two ``photon-decay-leptons.''
This part of the shower contains mostly electrons and positrons,
and the energy per particle will rapidly fall below the threshold for photon decay. In the second stage,
the particle multiplication proceeds in the conventional manner.
The energy per particle
starts at the Lorentz-violating photon-decay threshold energy
and drops until it is below the critical value $E_c$,
at which moment the particle number in the shower reaches its maximum.

We have found that negative $\ka$ could drastically reduce 
the shower-maximum depth $X_\text{max}$
of the electromagnetic shower as well as modify its energy dependence.
The absence of these features could be used to significantly improve the existing limits on negative values of $\ka$.

\section*{\hspace*{-5mm}ACKNOWLEDGMENTS}
\noindent
We thank B.~Altschul and D.~Kostunin
for useful comments  on an earlier version of this article.
This work was supported in part by the German Research Foundation (DFG)
under Grants No. KL 1103/4-1 and No. RI 1838/6-1.  

\begin{appendix}
\section{Heuristics of Lorentz-violating decays}
\label{app:Heuristics}

In this appendix, we briefly discuss the heuristics
of certain decay processes in the isotropic modified Maxwell
theory \eqref{L}--\eqref{k_F} based on the concept of the
effective mass-square~\cite{StrakhovOwen1996,ColemanGlashow1999,%
KaufholdKlinkhamer2005}. In particular, we make, as promised
in Sec.~\ref{subsec:Shower-model-with-LV},
some remarks on neutral-pion decay in the
Lorentz-violating (LV) theory considered, $\pi^0\to \LVga\,\LVga$.

From the dispersion relation implicit in \eqref{v_ph}
with $c=1$, we have the following definition of the
effective mass-square:
\beq\label{eq:disp-rel-effective-mass-square}
\omega^2= \frac{1-\ka}{1+\ka}\, |\vec{k}|^2
\sim
|\vec{k}|^2 + (-2\,\ka)\, |\vec{k}|^2
\equiv
|\vec{k}|^2+ m^2_{\text{eff},\,\LVga}(\vec{k})\,.
\eeq
In four-vector notation, this reads
\beq\label{eq:effective-mass-square}
k^2=
\omega^2 - |\vec{k}|^2
\sim
m^2_{\text{eff},\,\LVga}(\vec{k})
\equiv
 -2\,\ka\, |\vec{k}|^2\,.
\eeq
The 3-momentum dependence of the
effective mass-square in \eqref{eq:effective-mass-square}
makes clear that the theory is Lorentz violating
(in the following, this 3-momentum dependence is often kept implicit).
As discussed extensively in Ref.~\cite{KaufholdKlinkhamer2005},
the effective mass-square allows for a heuristic
discussion of LV decay processes in general.
Here, we consider three decay processes in
the isotropic modified Maxwell theory with negative $\ka$.

The first process is
photon decay into an electron-positron pair~\cite{KS2008},
$\LVga\to e^-+e^+$.
The threshold photon energy  from \eqref{omega_th}
becomes for small $|\ka|$
\beq\label{eq:omega_th-app}
\om_\text{th} \sim  2\,m\;\sqrt{\frac{1}{-2\,\ka}}\,,
\eeq
which reads as follows in terms of the effective mass from
\eqref{eq:effective-mass-square}:  
\beq\label{eq:omega_th-effective-mass-square-app}
m_{\text{eff},\,\LVga}^\text{(th)} \sim  2\,m \,.
\eeq
Hence, the LV photon decay is as if a massive scalar
at rest decays into two scalars of mass $m$.

The second process is Bremsstrahlung by an electron
or positron in the Coulomb field of a nucleus,
$e^\pm \; (+\,Z)\to e^\pm +\LVga \;(+\,Z)$.
Below  \eqref{combined-process} in the main text,
we have already mentioned that
the rate would be essentially the same as in the Lorentz-invariant
theory. Indeed, consider energy conservation (3-momentum
conservation is taken care of by the nucleus),
\beq
\sqrt{E^2+M_e^2}= \sqrt{E'^2+M_e^2}
+ \sqrt{|\vec{k}|^2+ m^2_{\text{eff},\,\LVga}}\,,
\eeq
with initial electron energy $E$,
final electron energy $E'$, and photon 3-momentum
$\vec{k}$. Considering ultrarelativistic initial and final
electrons and using the effective mass-square
from \eqref{eq:effective-mass-square}, we get
\beq
E \sim  E' + |\vec{k}|\,(1-\ka)\,.
\eeq
A typical process at ultrahigh energies can
have $E'=E/2$ with the photon taking the
remaining energy $E/2$, just as for the case
of the standard Lorentz-invariant theory
(see Fig.~14 on p.~170 in Ref.~\cite{DKM2014}).

The third process is neutral-pion decay
in the theory with the Standard Model action augmented
by a single LV  photonic term proportional to $\kappa$
(taken to be negative and to have a very small magnitude).  
The energies and 3-momenta of the process are denoted as follows:
\beq
\pi^0 (E,\,\vec{q})\to \LVga(\omega,\,\vec{k})+\LVga(\omega',\,\vec{k}')\,,
\eeq
with the on-shell energies $E(\vec{q})=\sqrt{|\vec{q}|^2+(m_{\pi^0})^2}$
and $\omega(\vec{k})\geq0$ from
\eqref{eq:disp-rel-effective-mass-square}.
In terms of the photon effective mass-square from
\eqref{eq:effective-mass-square}, we have the cutoff
condition
\beq\label{eq:pi0-decay-heuristics}
m_{\pi^0} \sim 2\,m_{\text{eff},\,\LVga}^\text{(cutoff)}
\sim   \sqrt{-2\,\ka}\;\, (2\,\omega_\text{cutoff})\,.
\eeq
This gives, upon identifying the pion energy
$E\sim 2\,\omega$, the following cutoff energy for neutral-pion decay:
\beq\label{eq:E-cutoff}
E_{\pi^0}^{\,(\text{cutoff})} \sim \frac{m_{\pi^0}}{\sqrt{-2\,\ka}}\,,
\eeq
which agrees with the kinematic result of Ref.~\cite{ColemanGlashow1999}.
The heuristics of the first part of
\eqref{eq:pi0-decay-heuristics} makes clear that
neutral-pion decay into two LV photons $\LVga$
can occur for pion energies below \eqref{eq:E-cutoff} but not above.

The general expression for the neutral-pion decay
parameter $\gamma$~\cite{KaufholdKlinkhamer2005}
is as follows
(see, e.g., Sec.~3.6 of Ref.~\cite{DeWitSmith1986}):
\beqa\label{eq:gamma}
\gamma_{\pi^0\to \LVga+\LVga}\,(\vec{q})
&=&
\frac{1}{2}\,\frac{1}{(2\pi)^2}\,
\int \frac{d^3k}{2\,\omega(\vec{k})}\,
\int \frac{d^3k'}{2\,\omega(\vec{k}')}\,
\delta^4(q-k-k')\;|\mathcal{M}|^2\,,
\eeqa
with the symmetry factor $1/2$ and
the matrix element $\mathcal{M}$. Observe that,
in a Lorentz-invariant theory, the right-hand side of
\eqref{eq:gamma}
is manifestly Lorentz invariant and $\gamma$ is the
decay constant, which is independent of $\vec{q}$.
Taking the effective pion-photon-photon interaction 
(from anomalous triangle diagrams with quarks) to be~\cite{DeWitSmith1986},
\beq
\mathcal{L}_\text{eff}=\alpha\,C\,\phi\,
\epsilon^{\alpha\beta\gamma\delta}\,F_{\alpha\beta}\,F_{\gamma\delta}\,,
\eeq
with coupling constant $C$ of mass dimension $-1$,
the phase-space integral of the LV decay parameter \eqref{eq:gamma}
gives
\beqa\label{eq:gamma-phase-space-result}
\hspace*{-10mm}
\gamma_{\pi^0\to \LVga+\LVga}\,(\vec{q})
&=&
\left\{
\begin{array}{ll}
\displaystyle{\frac{1}{2}\,\frac{1}{8\pi}\,
\left(\frac{1-\ka}{1+\ka}\right)^{3/2}       
}
\,
\left[ \big(8\,\alpha\, C\big)^2\,\frac{1}{2}\,
\big(m_{\pi^0}\big)^4  + \cdots \right]
& \;\text{for}\;  E(\vec{q})< E_{\pi^0}^{\,(\text{cutoff})},\\[4mm]
0
& \;\text{otherwise}\,,
\end{array}
\right.
\eeqa
with the cutoff energy from \eqref{eq:E-cutoff}.
The ellipsis in the large square bracket of \eqref{eq:gamma-phase-space-result}
is because  the result for $|\mathcal{M}|^2$
depends on the LV parameter $\ka$, as the calculation
involves, for example, the $k^2$ of the photons
(see, e.g., Secs.~4.5 and 7.1 of Ref.~\cite{DeWitSmith1986} for
an extensive discussion of the standard Lorentz-invariant
decay process). The complete result for \eqref{eq:gamma-phase-space-result}
will be given elsewhere~\cite{Klinkhamer2016}.  

\begin{figure}[p]   
\centering
\includegraphics[width=0.6\textwidth]{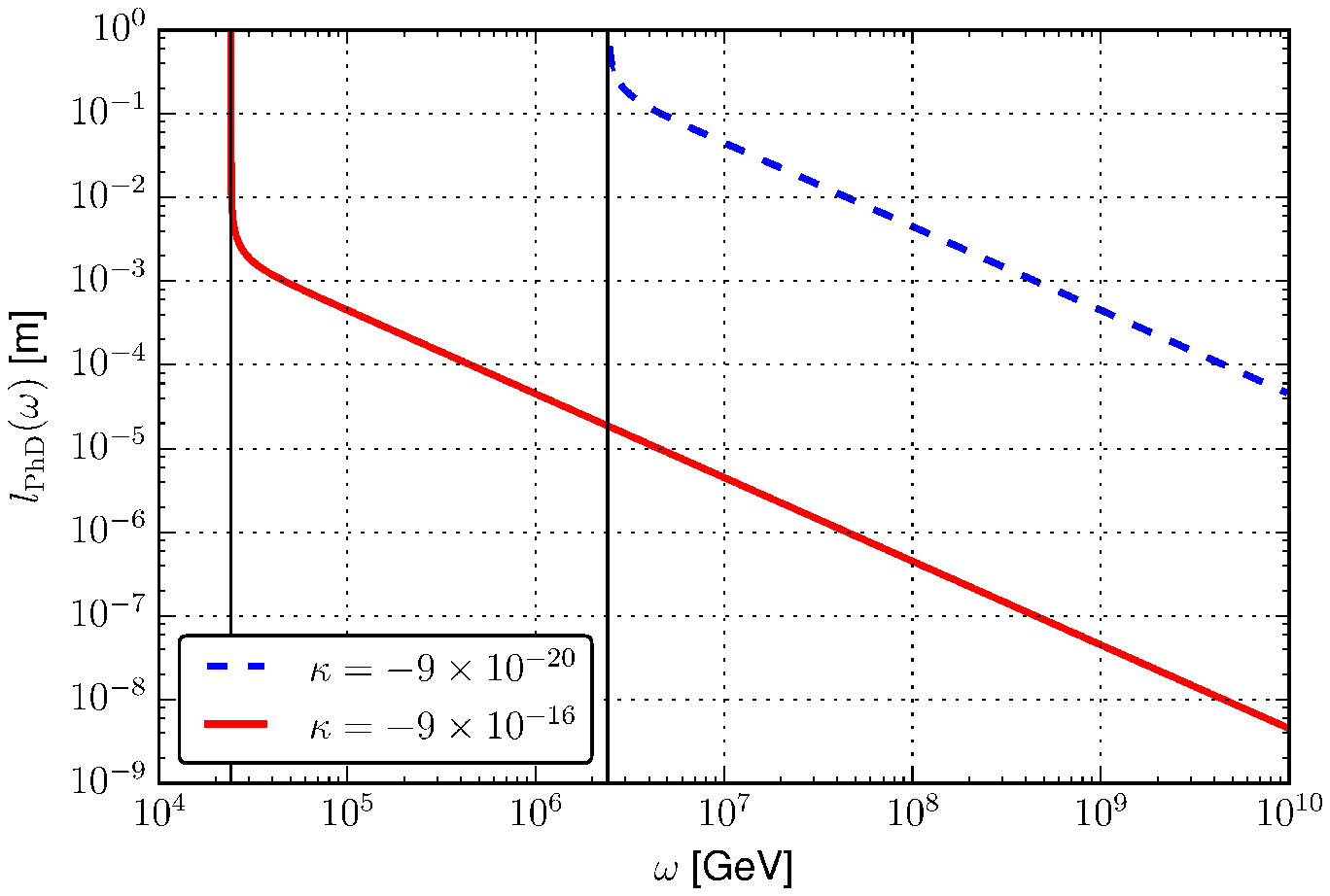}
\caption{Photon decay length $l_\text{PhD}\equiv 1/\Ga_\text{PhD}$ as a function of the photon energy $\om$ for the process $\LVga\to e^-+e^+$.
Two values of the Lorentz-violating parameter $\ka$ are considered,        
$\ka=-9\times10^{-16}$ and $\ka=-9\times10^{-20}$,
and  the vertical lines indicate the respective threshold energies,
defined by \eqref{omega_th} in terms of $\ka$ and the electron mass $m$.
For $\ka\uparrow 0$, the photon-decay threshold energy moves towards
infinity.
} \label{Fig:l_PhD}
\vspace*{5cm}
\centering           
\includegraphics[width=0.6\textwidth]{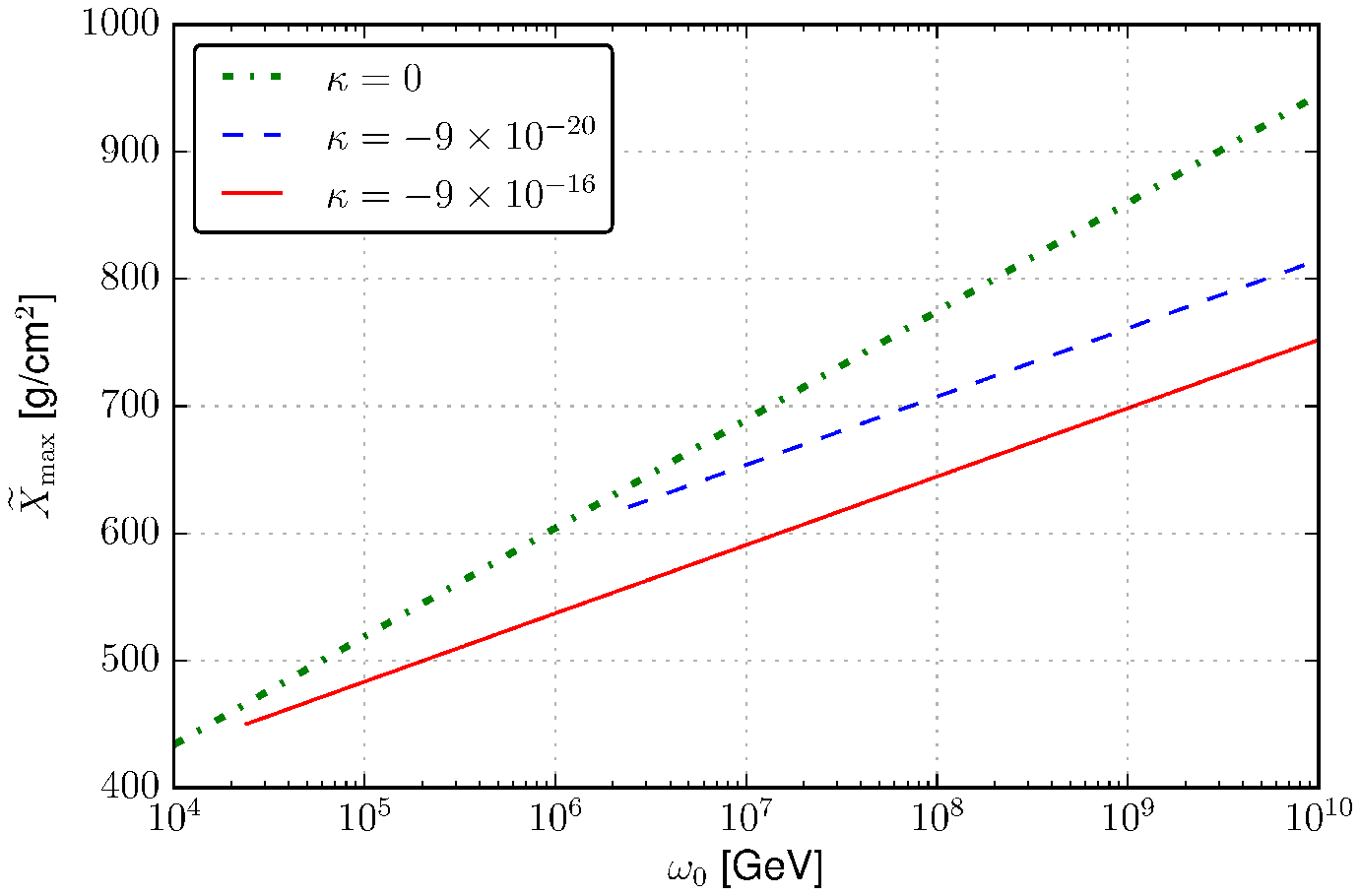}
\caption{Shower-maximum depth \eqref{widetilde-Xmax} of the electromagnetic shower for different values of $\ka$
as a function of the initial photon energy $\omega_0$. The blue-dashed curve for $\ka=-9\times10^{-20}$ has a larger photon-decay
threshold energy than the red-solid curve for $\ka=-9\times10^{-16}$.}
\label{Fig:Xmax(w)}
\end{figure}

\begin{figure}[p]   
\centering          
\includegraphics[width=0.6\textwidth]{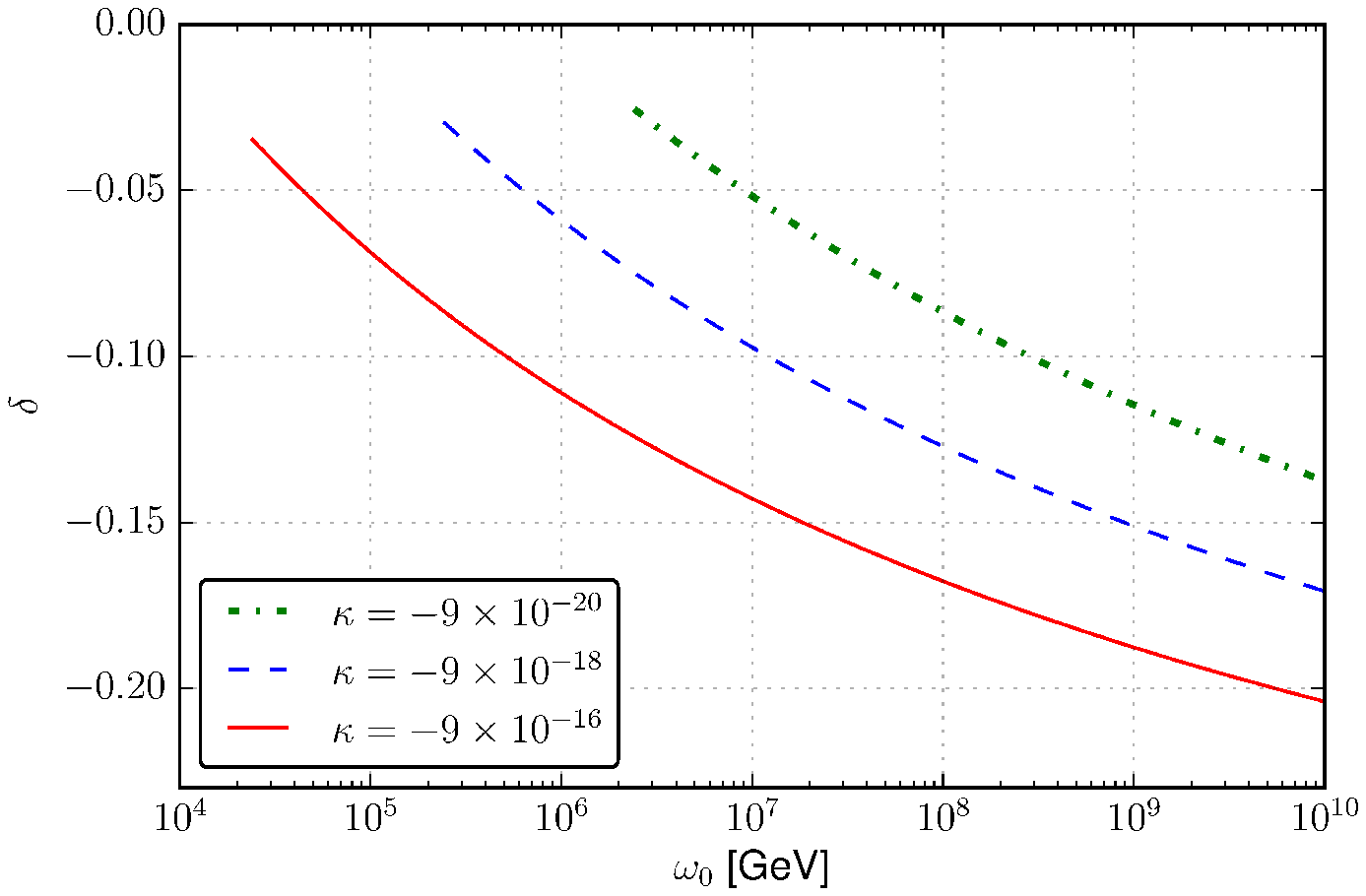}
\caption{Relative modification \eqref{delta}
of the shower-maximum depth \eqref{widetilde-Xmax-delta}
for different values of the Lorentz-violating
parameter $\ka$. If an electromagnetic shower is initiated by a $10^7$-GeV photon, a reduction of approximately 15\% on $X_\text{max}$
is expected from the current limit value $\ka=-9\times10^{-16}$,
whereas a 10\% reduction is expected from $\ka=-9\x10^{-18}$
and a reduction by 5\% from $\ka=-9\x10^{-20}$.
The three different curves start at different threshold energies defined
by \eqref{omega_th} in terms of $\ka$ and the electron mass $m$.}
\label{Fig:de(w)}
\end{figure}

\begin{figure}[p]
\centering                     
\includegraphics[width=0.6\textwidth]{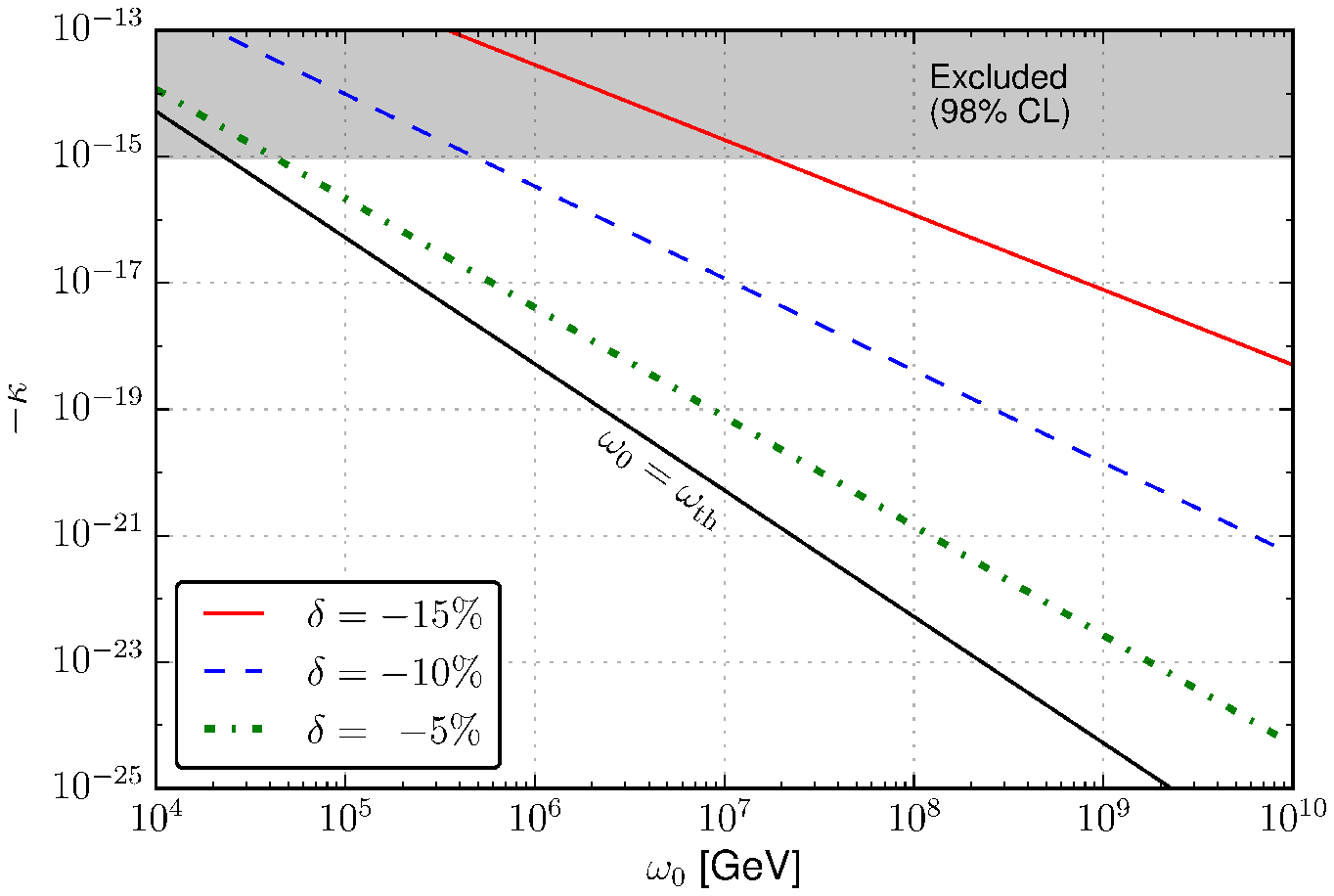}
\caption{Lorentz-violating parameter $\ka$ as a function of the
initial photon energy $\om_{0}$ producing different relative modifications $\delta$
of the shower-maximum depth.
Also shown is the line $\om_{0}=\om_\text{th}(\ka)$ with
$\om_\text{th}(\ka)$ given by \eqref{omega_th}
and only photons to the right of this line are considered.
If an electromagnetic shower is initiated by a $10^8$-GeV photon,
a reduction of 15\% on $X_\text{max}$
would be produced by $\ka\sim -10^{-16}$,
a reduction of 10\% by $\ka\sim-5\x10^{-19}$,
and a reduction of 5\% by $\ka\sim-2\x10^{-21}$. }
\label{Fig:k(w)}
\end{figure}

Regarding the neutral-pion-decay process, let us briefly turn to
extensive-air-shower phenomenology in our LV theory.
The neutral-pion-decay cutoff energy \eqref{eq:E-cutoff}
is about a factor 100 larger than the photon-decay threshold
energy \eqref{omega_th}, because of the different masses involved.
For a given negative value of $\ka$, there will be no
photons produced from neutral-pion decay with energies
$\omega_0 > 1/2 \times E_{\pi^0}^{\,(\text{cutoff})}$.
If we now look at Fig.~\ref{Fig:Xmax(w)} and translate
the initial photon energy $\omega_0$ on the horizontal axis
into a typical energy of a hadronic primary
(very roughly, $E_\text{prim} \sim 10 \times \omega_0$), then
the $\ka\ne 0$ curves for $X_\text{max}$ will, most likely,
receive an additional suppression for
$E_\text{prim} \gtrsim 10 \times E_{\pi^0}^{\,(\text{cutoff})}$.
These and other effects must be taken care of in a
numerical simulation, on which we plan to
report in a forthcoming publication.

\end{appendix}




\begin{thebibliography}{99}

\bibitem{KlinkhamerRisse1}
F.R.~Klinkhamer and M.~Risse,
``Ultra-high-energy cosmic-ray bounds on nonbirefringent modified-Maxwell theory,''
Phys.\ Rev.\ D {\bf 77}, 016002 (2008),
arXiv:0709.2502.

\bibitem{KlinkhamerRisse2}
F.R.~Klinkhamer and M.~Risse,
``Addendum: Ultrahigh-energy cosmic-ray bounds on nonbirefringent modified-Maxwell theory,''
Phys.\ Rev.\ D {\bf 77}, 117901 (2008),
arXiv:0806.4351.

\bibitem{KS2008}
F.R.~Klinkhamer and M.~Schreck,
``New two-sided bound on the isotropic Lorentz-violating parameter of modified Maxwell theory,''
Phys.\ Rev.\ D {\bf 78}, 085026 (2008),
arXiv:0809.3217.

\bibitem{DK2015}  
J.S.~D\'iaz and F.R.~Klinkhamer,
``Parton-model calculation of a nonstandard decay process in isotropic modified Maxwell theory,''
Phys.\ Rev.\ D {\bf 92}, 025007 (2015),
arXiv:1504.01324.

\bibitem{KM2007}
V.A.~Kosteleck\'y, and M.~Mewes,
``Lorentz-violating electrodynamics and the cosmic microwave background,''
Phys.\ Rev.\ Lett. {\bf 99}, 011601 (2007),
astro-ph/0702379.

\bibitem{DKM2014}
J.S.~D\'iaz, V.A.~Kosteleck\'y, and M.~Mewes,
``Testing relativity with high-energy astrophysical neutrinos,''
Phys.\ Rev.\ D {\bf 89}, 043005 (2014),
arXiv:1308.6344.

\bibitem{Diaz:2016c}
J.S.~D\'iaz and T.~Schwetz,
``Limits on CPT violation from solar neutrinos,''
Phys.\ Rev.\ D {\bf 93}, 093004 (2016),
arXiv:1603.04468.

\bibitem{BaileyKostelecky2004}
Q.~Bailey and V.A.~Kosteleck\'y,
``Lorentz-violating electrostatics and magnetostatics,''
Phys.\ Rev.\ D {\bf 70}, 076006 (2004),
arXiv:hep-ph/0407252.

\bibitem{Noordmans-etal2014}
J.P.~Noordmans, C.J.G.~Onderwater, H.W.~Wilschut, and R.G.E.~Timmermans,
``Question of Lorentz violation in muon decay,''
Phys.\ Rev.\ D {\bf 93}, 116001 (2016),
arXiv:1412.3257.


\bibitem{Antonov-etal2001}
E.E.~Antonov  {\it et al.},
``Test of Lorentz invariance through observation of the longitudinal development of ultrahigh-energy extensive air showers,''
JETP Lett.\  {\bf 73}, 446 (2001).


\bibitem{Altschul2007}
B.~Altschul,
``Astrophysical limits on Lorentz violation for pions,''
Phys.\ Rev.\ D {\bf 77}, 105018 (2008),
arXiv:0712.1579.

\bibitem{Boncioli-etal2015}
D.~Boncioli {\it et al.},
``Future prospects of testing Lorentz invariance with UHECRs,''
in: \textit{Proceedings of the 34th International Cosmic Ray Conference},
PoS ICRC2015 (2015),
arXiv:1509.01046. 


\bibitem{Diaz2014}
J.S.~D\'iaz,
``Neutrinos as probes of Lorentz invariance,''
Adv.\ High Energy Phys.\  {\bf 2014}, 962410 (2014),
arXiv:1406.6838.

\bibitem{Heitler1944}
W.~Heitler, \textit{Quantum Theory of Radiation},
second edition (Oxford University Press, London, 1944).

\bibitem{ChadhaNielsen1983}
S. Chadha and H.B. Nielsen,
``Lorentz invariance as a low-energy phenomenon,''
Nucl.\ Phys.\  B {\bf 217}, 125 (1983).

\bibitem{KM2002}
V.A. Kosteleck\'{y} and M. Mewes,
``Signals for Lorentz violation in electrodynamics,''
Phys.\ Rev.\  D {\bf 66}, 056005 (2002), arXiv:hep-ph/0205211.

 \bibitem{Hohensee2009}
M.A.~Hohensee, R.~Lehnert, D.F.~Phillips, and R.L.~Walsworth,
``Particle-accelerator constraints on isotropic modifications of the speed of light,''
Phys.\ Rev.\ Lett.\  {\bf 102}, 170402 (2009),
arXiv:0904.2031.


\bibitem{Altschul2009}
B.~Altschul,
``Bounding isotropic Lorentz violation using synchrotron losses at LEP,''
Phys.\ Rev.\ D {\bf 80}, 091901 (2009),
arXiv:0905.4346.

\bibitem{Altschul2005}
B.~Altschul,
``Lorentz violation and synchrotron radiation,''
Phys.\ Rev.\ D {\bf 72}, 085003 (2005),
arXiv:hep-th/0507258.


\bibitem{UHECRreview2011}
A.~Letessier-Selvon and T.~Stanev,
``Ultrahigh energy cosmic rays,''
Rev.\ Mod.\ Phys. {\bf 83}, 907 (2011),
arXiv:1103.0031.

\bibitem{Auger2015}
 A.~Aab {\it et al.} [Pierre Auger Collaboration],
``The Pierre Auger Cosmic Ray Observatory,''
Nucl. Instrum. Meth. A {\bf 798}, 172 (2015),
arXiv:1502.01323. 


\bibitem{Auger2014}
A.~Aab \textit{et al.} [Pierre Auger Collaboration],
``Depth of maximum of air-shower profiles at the Pierre Auger
Observatory. I. Measurements at energies above $10^{17.8}$ eV,''
Phys.\ Rev.\ D {\bf 90},  122005  (2014),
arXiv:1409.4809.  


\bibitem{StrakhovOwen1996} 
E.~Strakhov and D.A.~Owen,
``Decay of a Chern-Simons `photon' and $e^+$ $e^-$ MeV peaks,''
J.\ Phys.\ G {\bf 22}, 473 (1996).



\bibitem{ColemanGlashow1999}
S.R.~Coleman and S.L.~Glashow,
``High-energy tests of Lorentz invariance,''
Phys.\ Rev.\ D {\bf 59}, 116008 (1999),
arxiv:hep-ph/9812418.

\bibitem{KaufholdKlinkhamer2005}
C.~Kaufhold and F.~R.~Klinkhamer,
``Vacuum Cherenkov radiation and photon triple-splitting in a Lorentz-noninvariant extension of quantum electrodynamics,''
 Nucl.\ Phys.\ B {\bf 734}, 1 (2006),
arXiv:hep-th/0508074.


\bibitem{DeWitSmith1986}
B.~de~Wit and J.~Smith,
\textit{Field Theory in Particle Physics, Volume 1}
(North-Holland Publ., Amsterdam, 1986).

\bibitem{Klinkhamer2016} 
 F.R.~Klinkhamer,
``Lorentz-violating neutral-pion decays in isotropic modified Maxwell theory,''
arXiv:1610.03315.  

\end{thebibliography}
\end{document}